\documentstyle[12pt]{article}
\begin{document}
\begin{center}

~

\vskip1.0cm
\noindent
{\LARGE \bf Search for energetic cosmic axions utilizing
terrestrial/celestial magnetic fields}

\vskip1.0cm
{\sf K.~ZIOUTAS }

\vskip0.4cm
         
\noindent
{\it Physics Department, University of Thessaloniki, 
  GR-54006  Thessaloniki,  Greece.}

\vskip0.5cm 
{\sf D. J.~THOMPSON }

\vskip0.4cm

\noindent
{\it NASA/Goddard Space Flight Center, Code 660, Greenbelt MD 20771, USA.}

\vskip0.5cm
\small{\sf AND}

\vskip0.5cm
{\sf E. A.~PASCHOS }

\vskip0.4cm

\noindent
{\it Institut f\"ur Physik, Universit\"at Dortmund,
D-44221 Dortmund, Germany.}

\vskip1.0cm
{\it 12. 8. 1998 }

\vskip0.6cm
\end{center}

\noindent
{\bf Abstract.}
{\sf Orbiting $\gamma$-detectors combined with the magnetic field of the
Earth or the Sun can work parasitically as cosmic axion telescopes.
The relatively short field lengths allow the axion-to-photon conversion to be
coherent for $m_{axion} \sim 10^{-4}$ eV, if the  axion kinetic
energy is above $\sim 500$ keV (Earth's field), or, $\sim 50$ MeV
(Sun's field), allowing thus to search for axions from $e^+e^-$ annihilations,
from supernova explosions, etc.
With a detector angular resolution of $\sim 1^o$, a more efficient sky survey
for energetic cosmic axions passing {\it through the Sun} can be performed.
Axions or other axion-like particles might be created by the interaction
of the cosmic radiation with the Sun, similarly to the axion searches in
accelerator beam dump experiments; the enormous  cosmic energy combined
with the built-in coherent Primakoff effect might provide a sensitive 
detection scheme, being out of reach with accelerators.
The axion signal will be an excess in $\gamma$-rays coming either from a
specific celestial place behind the Sun, e.g. the Galactic Center, or,
from any other direction in the sky being associated with a violent
astrophysical event, e.g. a supernova.
Earth bound detectors  are also of potential interest.
The axion scenario also applies to
other stars or binary systems in the Universe, in particular to those with
superstrong magnetic fields.}

\vskip1.8cm
\noindent
{\sf ----------------------------------------------------------

\noindent
{\it Emails} :  Konstantin.Zioutas@cern.ch

\noindent
~~~~~~~~~~~~DJT@bartok.gsfc.nasa.gov

\noindent
~~~~~~~~~~~~Paschos@hal1.physik.uni-dortmund.de   }

\newpage

\section*{1. Introduction}

An attractive solution of the strong CP problem invokes a new symmetry,
the Peccei-Quinn (PQ) symmetry (U$_{PQ}$(1)).
The spontaneous breaking of this new symmetry predicts the existence of a
light neutral pseudoscalar particle, the {\sf \it axion}, which is closely
related to the neutral pion
\cite{pq,ww}.
In fact, there are good reasons to believe that if the PQ mechanism is
responsible for preserving $CP$ in the strong interactions, then the
{\sf \it axion} is the dark matter
\cite{kamion},
i.e. {\sf \it axions} may exist as primordial cosmic relics copiously 
produced in the early Universe, and eventually thermalized in a way similar
to the 2.7$^o$K cosmic background radiation.
The {\sf \it axion} also arises in supersymmetry and superstring theories.
Thus, the {\sf \it axion} is one of the leading and promising
non-baryonic candidate for the ubiquitous dark matter in the universe
\cite{ggr}.
Astrophysical observations and laboratory experiments leave open an
{\sf \it axion} rest mass window around  $m_a \approx 10^{-4}~$eV
(within $\pm$1-2 orders of magnitude).
For all these reasons, {\sf \it axions} have
received much attention in elementary particle (astro)physics.

The {\sf \it axion} decay into two photons
(${\sf \it a \rightarrow \gamma \gamma}$) is the reaction mainly used to
search for them. Inside matter or in a magnetic field in vacuum,
the {\sf \it axion} couples to a virtual photon (Primakoff effect),
producing  a real photon ($\gamma$),
which can be  detected :
\begin{equation}
a ~ +~ \gamma_{virtual} \rightleftharpoons \gamma 
\end{equation}
The {\sf \it axion} behaves like a very weakly interacting photon or pion
($\pi^0$), and,
in a reaction, it can replace a magnetic dipole $\gamma$ or a $\pi^0$.
Energetic {\sf \it axions} with mean thermal energy
equal to $\sim$4 keV
\cite{dml,zzz}
or $\sim$ 160 MeV
\cite{emrt}
are possibly copiously produced via the Primakoff effect inside the Sun
or during a Supernova explosion, respectively.
They also could be produced in astrophysical beam dumps
\cite{gaisser},
similar to the beam dump in accelerators, replacing energetic $\pi^0$'s or
$\gamma$'s in the electromagnetic/hadronic cascade reactions involved.
Therefore, energetic {\sf \it axions}
have been searched already in high energy neutrino experiments
\cite{oldneu}.
A not so unusual beaming effect can compensate for the large distance to the
Earth.  In other words, it is not so unreasonable to expect high energy
cosmic {\sf \it axions}, beyond those  expected to be emitted from a
supernova.
Finally, the existence of  particles beyond the Standard
Model with similar couplings can not be excluded.

\section*{2. Previous work}

The stimulation for this proposal comes from two recent  works, which
appeared almost simultaneously by two groups
\cite{emrt}
searching for energetic {\sf \it axions} from SN1987A with  data
from orbiting detectors. However, they could have provided an {\sf \it axion}
signature for specific parameter values, i.e., for $m_a \leq 10^{-9}$eV and
an {\sf \it axion}-to-photon coupling constant
$g_{a\gamma \gamma} \geq 3\cdot 10^{-12}$ GeV$^{-1}$ assuming
$\sim 1~kpc~=~3\cdot 10^{19}m$
galactic magnetic field length of $\sim 2~\mu gauss$, where the
{\sf \it axion}-to-photon conversion via the coherent Primakoff effect
can take place.
In such a case, an orbiting detector pointing to the SN1987A position in the
sky should have measured an excess of energetic photons during this
historical supernova observation on Earth, provided the number of the emitted
{\sf \it axions} were sufficient to trigger the detector.
These two groups
\cite{emrt}
have also estimated the hypothetical thermal {\sf \it axion} spectrum from
the supernova, with the main unknown being the
{\sf \it axion}-to-photon coupling constant ($g_{a\gamma \gamma}$).

\section*{3. The suggestion}

The concept of this work is similar with that given in ref.
\cite{emrt}.
The main difference is the choise of the magnetic field
between  the {\sf \it axion} source and the $\gamma-$ray detector,
where the {\sf \it axion} conversion takes place;
we suggest to use the solar ($\sim 2 ~gauss$) and/or the terrestrial
($\sim 0.5 ~gauss$) magnetic fields.
In order to justify this choise, we give below two relations, which
describe the  {\sf \it axion} interaction inside a static magnetic field.

Firstly, the {\sf transverse} magnetic field strength {\sf B}, its length
{\sf L} and the {\sf \it axion}-to-photon coupling constant
$g_{a\gamma \gamma}$ are the fundamental parameters in the calculation of
the coherent {\sf \it axion}-to-photon conversion inside {\sf B}.
The probability to have a photon from an energetic {\sf \it axion}
entering perpendicularly a 1 T$\cdot$m magnetic field ,
is
\cite{dml,zzz,grls} 
\begin{equation}
P_{a\rightarrow \gamma} = \left(\frac{gBL}{2}\right)^2=
2.5\times 10^{-21} \left[\frac{ {\sf B}}{1T}\frac{{\sf L}}{1m}
\frac{{\sf g_{a\gamma \gamma}}}{10^{-10}GeV^{-1}}\right]^2 ,
\end{equation} 
It is interesting to notice the
{\sf $(B\cdot L)^2$} dependence of the coherent {\sf \it axion}-to-photon
conversion rate.

Secondly, for massive {\sf \it axions}, in order to fullfil the coherence
relation (2), i.e. to exclude deconstructive {\sf \it axion}-photon
interference over the magnetic field length ({\sf L}), the {\sf \it axion} rest
mass ($m_a$) and its total energy ($E_a = E_\gamma =\hbar \omega$)
must satisfy a second relation
\cite{dml,zzz}
\begin{equation}
L~ \leq~ \frac{(2\pi \hbar c)\cdot (\hbar \omega)}
{|m_a^2-m_{\gamma}^2|c^4}
\end{equation}
In this relation, $m_{\gamma}$ reflects the acquired mass of the photons
inside matter, which depends on the electron density :
$m_{\gamma}[\mu eV]\approx 0.37\times \sqrt{\rho_e[10^8/cm^3]}$.
Thus, for an electron density $\rho_e \leq 10^8/cm^3$
(i.e. $m_{\gamma}\ll m_a$), which can be the case with
the considered terrestrial and solar regions
\cite{allen},
relation (3) becomes
\begin{equation}
L~ \leq~ \frac{(2\pi \hbar c)\cdot (\hbar \omega)}
{m_a^2c^4}
\end{equation}
Inserting  $m_a\approx 10^{-4}$eV and $\bar{E_a}\approx 160$ MeV into this
relation, the resulting coherence length (for $E_a \geq 50$ MeV) can be as
large as
\begin{equation}
L ~\approx ~ 4\cdot 10^{9}~m ~\approx~ 6R_{\odot}
\end{equation}
Similarly, for $m_a\approx 10^{-4}$eV and  $E_a~=~511~keV$ it follows :  
\begin{equation}
L ~\approx ~ 10^{7}~m ~\approx~ 1R_{\oplus}
\end{equation}
One should notice that the study in ref.
\cite{emrt}
is sensitive to {\sf \it axions} with rest mass below $10^{-9}$eV, because
they used the much longer coherent-galactic-magnetic field ($\sim 1~kpc$);
this mass range  is far below the open {\sf \it axion} mass window
($m_a \approx 10^{-(4\pm 2)}$ eV).

Thus, taken into account the coherence lengths given in relations (5) and (6),
the Earth's magnetic field is in this respect just appropriate for an
{\sf \it axion} threshold energy of $\sim$ 500 keV, while the more efficient
solar magnetic field, fits to high energy {\sf \it axions}
($E_a \geq 50$ MeV). We take for the terrestrial and solar  $B\cdot L$
values
\begin{equation}
(B\cdot L)_{\oplus} ~\approx~ 300~T\cdot m, ~~~and,~~~
(B\cdot L)_{\odot}~\approx ~3\cdot 10^4~T\cdot m ~,
\end{equation}
respectively. It is worth mentioning that solar flares with $\sim kgauss$
magnetic fields and some  $10^3~ km$ in size can have
$B\cdot L \approx 10^5-10^6~ T\cdot m$.
For comparison, one should keep in mind that laboratory
magnetic {\sf \it axion} spectrometers use magnetic fields with
$B\cdot L \approx 10-100~T\cdot m$ at best
\cite{dml,zzz,shigetaka};
inspite of the $\sim 2-10$ T field
strength, they have to be anyhow short (s. relation (4)), because of the much
lower expected solar {\sf \it axion} energy ($\sim$ 4 keV).

With relation (2) we can estimate the conversion probability
$P_{a\rightarrow \gamma}$ for an energetic {\sf \it axion} propagating inside
the Sun's, or, Earth's magnetic field :
\begin{equation}
P_{a\rightarrow \gamma}^{\odot} = 
2.5\times 10^{-12}\cdot \left[\frac{ {\sf (B\cdot L)}}{(3\cdot 10^{4}~T\cdot m)}
\frac{{\sf g_{a\gamma \gamma}}}{10^{-10}GeV^{-1}}\right]^2 
\end{equation} 
and
\begin{equation}
P_{a\rightarrow \gamma}^{\oplus} = 
2.5\times 10^{-16}\cdot \left[\frac{ {\sf (B\cdot L)}}{(300~T\cdot m)}
\frac{{\sf g_{a\gamma \gamma}}}{10^{-10}GeV^{-1}}\right]^2 
\end{equation} 
The Earth's magnetic field allows in principle for a simultaneous
$\sim 2\pi$ survey of the sky, even though the field of view (f.o.v.) of an
individual (orbiting) detector is smaller.
This is not the case with the solar field; as the Earth and the
Sun change continuously their orientation in space, one can scan
`{\it through the sun}' a big part
of the sky with $\sim 0.5^o$ opening angle.
We consider an effective solar {\sf \it axion} conversion region of a few
solar radii (see relation (5)) including the Sun.

To be more specific, we give a numerical example for a
{\sf supernova}, which might be taken as reference for other violent
astrophysical events. The expected integrated {\sf \it axion} flux ($\Phi_a$)
on Earth from a supernova explosion, which lasts some $\sim 10$-20 seconds
and is at a distance $D\approx6~kpc$, is
\cite{emrt} 
\begin{equation}
\Phi_a(\bar{E_a}\approx 160~MeV)~\approx~2.5\cdot 10^9~axions\cdot cm^{-2}\cdot \left[
\frac{g_{a\gamma \gamma}}{10^{-10}GeV^{-1}}
\right]^2 \cdot \left[\frac{6~kpc}{D}\right]^2,
\end{equation}
with the {\sf \it axions} created from the scattering of thermal photons on
protons through the Primakoff effect ({\it $p\gamma \rightarrow pa$}).
The energy released by the {\sf \it axions} is $\sim 100$ times smaller
than the energy released through the escaping neutrino burst.

Combining relations (8), (9) and (10), we estimate the signal
$S^{\odot} = P_{a\rightarrow \gamma}^{\odot}\times \Phi_a$
for an orbiting detector.  The flux of {\sf \it axions} from the supernova
converts into photons of $\sim$160 MeV in the solar field and gives the final
signal :
\begin{equation}
S^{\odot}~\approx~6\cdot 10^{-3}~cm^{-2}\cdot \left[\frac{g_{a\gamma \gamma}}{10^{-10}
GeV^{-1}}\right]^4
~\approx~8~cm^{-2}\cdot \left[\frac{g_{a\gamma \gamma}}{6\cdot 10^{-10}
GeV^{-1}}\right]^4
\end{equation}
Similarly, the intervening terrestrial field yields :  
\begin{equation}
S^{\oplus}~\approx~6\cdot 10^{-7}~cm^{-2}\cdot \left[\frac{g_{a\gamma \gamma}}{10^{-10}
GeV^{-1}}\right]^4
~\approx~8\cdot 10^{-4}~cm^{-2}\cdot \left[\frac{g_{a\gamma \gamma}}{6\cdot 10^{-10}
GeV^{-1}}\right]^4
\end{equation}
Notice the $(g_{a\gamma \gamma})^4$ dependence of the signal, while
$g_{a\gamma \gamma} \leq 6\cdot 10^{-10} GeV^{-1}$ is the presently best
experimental limit for the coupling constant
\cite{shigetaka}.

In a supernova explosion electrons and {\sf positrons} are created by the
interacting neutrinos above the neutrinosphere ($\rho\leq 10^{11} g/cm^3$)
via the reactions
$\nu \bar{\nu} \rightarrow e^+e^-,~ \nu_e p\rightarrow pe^- $ and
$\bar{\nu_e} p\rightarrow ne^+$, the dominant processes which generate
neutrino opacity
\cite{rcd}.
A similar situation might arise in the case of two merging neutron stars
\cite{eichler}.
While the annihilation $\gamma$-line of those positrons is completely
shielded, {\sf \it axions} created in the $e^+e^-$-annihilation can escape.
It is worth remembering that 511 keV {\sf \it axion} search, following the
reaction $e^+e^-\rightarrow \gamma a$, has been performed already in
laboratory experiments with positron sources
\cite{carboni}.
Needless to say that the same process could occur with those obscured intense
positron places in the Universe, while the intervening terrestrial or any
other magnetic field works as {\sf \it axion}-to-photon converter
(s. section 5.).

Obviously, the solar signature ($S^{\odot}$) will show-up when the detector,
the Sun and the {\sf \it axion} source are aligned within $\sim 0.5^o$,
in which case the f.o.v. of the detector sees the {\sf \it axion} source. The
geometry with the Earth's magnetic field is actually free from such
constraints, however, the conversion efficiency is smaller (s. relation (8)
and (9)). In other words, taking into account the suggested {\sf \it axion}
conversion inside the terrestrial or solar magnetic field in the evaluation
of $\gamma$-ray data, an orbiting gamma detector is also an {\sf \it axion}
telescope, scaning continuously the sky for such events.

\noindent
{\bf Background:}
The isotropic cosmic $\gamma$-ray flux above $\sim${\sf 100 MeV} is
$\sim 10^{-5} \gamma ~'s/cm^2\cdot s\cdot sr$
while that from the Galactic plane is by a factor of  $\sim 10$ higher
\cite{moon}.
Furthermore, the electromagnetic/hadronic interaction of the cosmic radiation
with the Sun must give rise to energetic photons, and in addition there are
$\gamma$-rays from solar activity, e.g. flares
\cite{flares}.
Sofar, the orbiting EGRET detector
measured an excess of high-energy gamma radiation from the moon.
The lunar flux above 100 MeV is
$\sim 5\cdot 10^{-7} \gamma ~'s/cm^2\cdot s$,
while the limit obtained for the quiet Sun is below
$\sim 2\cdot 10^{-7} \gamma ~'s/cm^2\cdot s$
\cite{moon}.
The {\sf 511 keV} flux from the Galactic Center is
$\sim 2\cdot 10^{-4} \gamma ~'s/cm^2\cdot s$, with the $3\gamma$ annihilation
continuum below 511 keV being by a factor $\sim 5$ higher 
\cite{purcell}.
The observed Galactic positron annihilation rate is $\geq 10^{43}/sec$
\cite{rr};
those positrons can be created through the decay from radioactive nuclei
produced by supernovae, novae, and the massive Wolf-Rayet stars with violent
surface activity, but also from $\gamma$-$\gamma$ pair production in the
vicinity of an accreting black hole
\cite{purcell},
whose violence is probably without analog.

\noindent
{\bf The axion signature} : {\sf a)}~The {\sf \it axion} signal associated
with the Earth's field will be burst-like and in coincidence with some violent
astrophysical event, e.g. a supernova. This is a type of signal discussed in
ref.
\cite{emrt}.
~{\sf b)}~Cosmic {\sf \it axions}
converted in the solar field can be identified as an excess of $\gamma$-rays
coming from the region of the Sun, provided the direction from the detector
to the Sun points at the same time to a specific source outside the Sun,
e.g. the Galactic Center.
If a $\gamma$-ray excess is seen coincident with a
radio/optical/x-ray flare, then the gamma radiation could be from the flare.
These flares are monitored continuously by the GOES satellite, so screening
out solar flare events is straightforward.
\footnote{The rather strong magnetic fields in solar flares are also
quite suitable {\sf \it axion} converters.
With better understanding of solar flares the question arises whether
$\gamma$-rays from {\sf \it axions} could be identified within solar flares
observations (s. ref. \cite{cc} for very low-mass axions in the x-ray region).}

\section*{4. Orbiting and Earth bound detectors}

The requirements for the high energy gamma detector are actually obvious from
the previous considerations of the potential cosmic {\sf \it axion}
signature. We give here in summary  the main
optimum specifications :
1) threshold energy $\sim 0.5/50~MeV$, with a modest energy
resolution being sufficient, 2) background suppression,
3) large detector aperture/f.o.v.,
4) angular resolution of a few degrees and 5) photon identification.

For example, EGRET satisfies most of the above requirements.
The planned GLAST project
\cite{glast1},
with an effective area $\sim 8000~cm^2$ (above 1 GeV),
f.o.v. covering $\sim 20\%$ of the sky, angular resolution of
$2.5^o $ (at 100 MeV) and $0.1^o$ (above 10 GeV),
energy range 10 MeV to 300 GeV, energy resolution $\sim 10\%$, will be
a factor of $\sim 30$ more sensitive than EGRET and it can become the best
potential high energy cosmic {\sf \it axion} antenna in orbit.

For the search of the 511 keV line different orbiting detectors come into
question. The OSSE detector is certainly the best instrument in orbit since
1991
\cite{osse};
it has $\sim 2000~cm^2$ aperture at 511 keV, f.o.v.
$3.8^o\times 11.4^o$, an energy resolution of 8$\%$ at 661 keV, while its
energy range from 50 keV to 10 MeV is just complementary to the GLAST
performance. The planned  European mission INTEGRAL will also be sensitive
to 511 keV {\sf \it axions} (energy range 15 keV to 10 MeV);
its f.o.v. will be $4.8^o-16^o$ and its targets of observation will include
the Galactic Center. Further, one should reconsider data from detectors,
which have had within their f.o.v. the region of SN1987A, eventhough there
is as yet no {\sf \it axion} flux  estimate at 511 keV from astrophysical
places like a supernova or other source in the sky.

Following the same reasoning at high energies, {\sf Earth bound detectors}
come also into question, provided they have the required photon
identification signature, and, the directional reconstruction of the
incident photon. However, a sky survey `through the Sun' requires a solar
blind $\gamma$-detector, i.e., the detection technique can not use
atmospheric Cherenkov  or scintillation light in the visible.
The high energy $\gamma$-radiation seen from the Moon with EGRET
\cite{moon}
and the observed shadowing of cosmic rays by the Sun and the Moon,
with surface
\cite{surface}
and deep underground detectors
\cite{ambrosio},
show the feasibility of this kind of investigations.
\footnote{Similarly, in accelerators, high energy detector
systems  with their inner charged particle tracking magnetic field
surrounded by $\sim 4\pi$
electromagnetic calorimeter can also perform this kind of investigations.
In fact, they can operate parasitically for this purpose, provided  they
have a built-in trigger, which allows to register any cosmic ray hitting
the detector when the accelerator is OFF, or between  beam crossings.
The potential {\sf \it axion} signature, i.e. isolated energetic photons
coming from the magnetic field region, will be of interest independent on the
time of occurence or direction of arrival.
Fortunately, background measurements can be performed by
switching OFF the magnetic field. The effective $B\cdot L$ value is usually
$\approx 1-10$ $T\cdot m$, while the $\sim 1~m$ transverse field length makes
them  coherent high energy {\sf \it axion} converter for an {\sf \it axion}
rest mass up to 1-10 eV;
their geometry allows to perform, however, {\sf simultaneously} a full sky
high-energy {\sf \it axion}
survey. To the best of our knowledge, none magnetic detector was ON to
convert energetic or $\sim$511 keV {\sf \it axions} during SN1987A.

A search for energetic {\sf \it axions} can also be performed with the
powerful accelerator bending magnets
\cite{zzz},
which have $B\cdot L \approx 100~T\cdot m\approx (B\cdot L)_{\oplus}$ and
a built-in angular resolution/acceptance of $\leq 0.5^o$.}

\section*{5. Discussion}

We have used a supernova explosion as a representative astrophysical violent
event, for which a possible  {\sf \it axion} involvement below $\sim 300$ MeV
has been estimated already quantitatively. However, it is reasonable to assume
that if {\sf \it axions} or any other {\sf \it axion}-like particles exist,
then, they could be copiously created in other flare stars or in transient
brightenings, which we know to release comparable or greater energy.
This work suggests primarily to utilize the terrestrial and the solar magnetic
field as {\sf \it axion}-to-photon converters, in order to perform with
orbiting  detectors a sky survey, searching for cosmic {\sf \it axions} with
energy above $\sim 0.5/50$ MeV.

An {\sf \it axion} signature can show-up either as a burst, or as an event
rate being proportional to the intervening $(B\cdot L)^2$ value between
the hypothetical {\sf \it axion} source and the detector.
This can be the case, for example, with the Earth's field by comparing
gamma rates observed at different distances from the Earth. For example,
the INTEGRAL mission will fly between $\sim 10000~km$ and $\sim 150000~km$.

We also mention  a few other places in the sky as potential sources for
{\sf \it axions}.

\noindent
{\bf a)}
{\sf astrophysical `beam dumps'}
\cite{gaisser},
e.g. relativistic `fireballs', jets, etc.,
which seem to be  associated with the as yet enigmatic Gamma Ray Bursts
(GRBs); the most powerful explosions in the Universe after the Big Bang :
the released energy (some $10^{52\pm 2}~ergs$) is probably much more than
that from a supernova explosion
\cite{BP,dar}.

\noindent
{\bf b)}
the {\sf Galactic  Center} (GC), which is one of the most dynamical regions
in our Galaxy, with numerous activities remaining hidden.
For example, EGRET observed a $\gamma$-ray source luminosity
$L\approx 10^4 L_{\odot}$ in the energy range 30 MeV to 20 GeV 
\cite{egret1}.
Further, the recent OSSE discovery of a giant cloud of positrons extending
$\sim 1~kpc$ above the GC was unexpected, since antimatter is thought to be
relatively rare in the Universe.

\noindent
{\bf c)}~{\sf close binaries}, e.g. cataclysmic variables, hypernova
\cite{BP},
etc.

Inspite of missing predictive theoretical models for energetic cosmic
{\sf \it axions} beyond those to be emitted from a supernova,
we propose to implement this kind of investigations in the different photon
detectors in orbit or on Earth. The realization of these investigations
require only an appropriate data re-evaluation and/or trigger.
Such sky surveys might unravel novel physical processes occuring deep inside
a star or our Galaxy.

For cosmic {\sf \it axions} with energy far above that expected to be emitted
from a supernova, i.e. $E_a\gg 10^8$eV, also the $\sim kpc$ galactic magnetic
fields considered in ref.
\cite{emrt}
can be very efficient {\sf \it axion} converters for an {\sf \it axion} rest
mass in the open {\sf \it axion} mass range (s. rel. (4)). Of course,
no orbiting detector can measure the energy of such  very energetic photons.
However, for the purpose of this suggestion, a mere photon identification
might well be sufficient as a first signature.

An additional perspective, of no minor importance, is the  possibility to
create and detect, during the quasi `beam dump' of the cosmic radiation into
the Sun, {\sf \it axions} or other new weakly interacting particles with
similar couplings
\cite{nomad1}.
Because of the huge thickness of the Sun, even a very weakly interacting
component of the cosmic radiation might interact there, which is beyond reach
in accelerator beam dump experiments. In addition, the advantage of this
configuration compared with accelerator experiments is the much higher
cosmic energy, combined with the built-in highly efficient coherent
{\sf \it axion}-to-photon conversion inside the solar magnetic field.

Finally, inspired by the celebrated microlensing phenomenon
\cite{BP0},
it does not escape our attention that the considered alignement between the
cosmic {\sf \it axion}-source, the solar field and the $\gamma$-detector can
also happen with another magnetic star in the sky replacing the Sun
\cite{ra}.
The {\sf \it axion} interaction  can be enhanced in stars having
strong magnetic fields, e.g. for $B\geq 10^{12}~gauss $
\cite{BP,magnetars}
around a neutron star, or $B\leq 10^9~gauss$ around a white dwarf
\cite{ra},
where the $B\cdot L$-values can be above $\sim 10^{12}~T\cdot m$;
for certain parameter values, the conversion efficiency
{\sf \it axion}-to-photon, and {\it vice versa}, might reach reasonable
values.
In particular, the {\sf \it axion} scenario could be present in eclipsing
(close) binaries with superstrong magnetic fields, whose configuration might
imply a high degree of alignment with the Earth.
The small size  of a neutron star allows coherence Primakoff  interaction
also in the x-ray region
\cite{morris}.
Therefore, if {\sf \it axions} exist, they could be responsible for some of
the time variable or transient cosmic $\gamma$-ray sources, including  GRBs
and  Soft Gamma-ray Repeaters.

\vskip1.2cm

\noindent
{\bf Acknowledgements}

\vskip0.4cm
\noindent
Two of us (E.A.P. and K.Z.) like to thank NATO for a cooperative research
grant.


\newpage



\begin{thebibliography}{99}

\bibitem{pq} R.D. Peccei, H.R. Quinn, Phys. Rev. {\bf D16} (1977);
 Phys. Rev. Lett. {\bf 38} (1977) 1440.

\bibitem{ww} F. Wilczek,  Phys. Rev. Lett. {\bf 40} (1978) 279;
 S. Weinberg, {\it ibid} {\bf 48} (1978) 223.
\bibitem{kamion}  M. Kamionkowski, CU-TP-866, CAL-648,
 hep-ph/9710467 (24. 10. 1997).

\bibitem{ggr} G. G. Raffelt, Proc. XVIII Intern. Conference on Neutrino
 Physics and Astrophysics, ed. Y. Suzuki and Y. Totsuka, Takayama, Japan
 (June 1998). See also hep-ph/9806506 (27.6.1998), and references therein.

\bibitem{dml} D. M. Lazarus, G. C. Smith, R. Cameron, A. C. Melissinos, G. Ruoso, 

 Y. K. Semertzidis, F. A. Nezrick,  Phys. Rev. Lett. {\bf 69} (1992) 2333.

\bibitem{zzz} K. Zioutas et al., astro-ph/9801176 (18. 1. 1998),
 N.I.M. A (1998) in press.

\bibitem{emrt} E. Masso and R. Toldra, Phys. Rev. {\bf D52} (1995) 1755,
 J. A. Grifols, E. Masso,

  R. Toldra, Phys. Rev. Lett. {\bf 77} (1996) 2372,
 E. Masso, astro-ph/9704056 and

 J. W. Brockway, E. D. Carlson, G. G. Raffelt,
 Phys. Lett. {\bf B383} (1996) 439.

\bibitem{gaisser} T. K. Gaisser, {\it Cosmic Rays and Particle Physics},
 Cambridge University Press (1990) 177.

\bibitem{oldneu} T. Hansl et al., Phys. Lett. {\bf 74B} (1978) 139,
 P. Alibran et al., {\it ibid} {\bf 74B} (1978) 134,
 P.C. Bosetti et al.,  {\it ibid} {\bf 74B} (1978) 143,
 F. Bergsma et al., {\it ibid} {\bf 157B} (1985) 458.

\bibitem{grls} G. Raffelt and L. Stodolsky, Phys. Rev. {\bf D37} (1988) 1237;

  G. Raffelt, Phys. Rev. {\bf D33} (1986) 897.

\bibitem{allen} C. W. Allen, {\it Astrophysical Quantities}, 2nd edition,
 University of London, The Athlone Press (1963) pp. 133, 176.

\bibitem{shigetaka} S. Moriyama, M. Minowa, T. Namba, Y. Inone, Y. Takusu,
 A. Yamamoto, RESCEU-23/98, hep-ex/9805026, submitt. to Phys. Lett. {\bf B}
  (1998).

\bibitem{rcd} R. C. Duncan, S. L. Shapiro, I. Wasserman, ApJ. {\bf 309}
 (1986) 141;

 S. E. Woosley, E. Baron, ApJ. {\bf 391} (1992) 228;

 S. E. Woosley, Astron.  $\&$ Astrophys. Suppl. Series {\bf 97} (1993) 205.

 

\bibitem{eichler} D. Eichler et al., Nature {\bf 340} (1989) 126.


\bibitem{moon} P. S. Sreekumar et al., ApJ. {\bf 494} (1998) 523;
 D. J. Thompson, D. L. Bertsch, D. J. Morris, R. Mukherjee,
 J. Geophys. Res. - Space Phys. {\bf 102}, A7 (1997) 14735;
 S. D. Hunter et al., ApJ. {\bf 481} (1997) 205.
 
\bibitem{flares} E.g. : G. Kanbach et al., Astron. $\&$ Astrophys. Suppl. Series
 {\bf 97} (1993) 349;

 N. G. Leikov et al., {\it ibid} {\bf 97} (1993) 345.

\bibitem{purcell} W. R. Purcell et al., ApJ. Lett. {\bf 417} (1993) 738,
 ApJ. {\bf 491} (1997) 725;

 R. L. Kinzer et al., Astron. $\&$ Astrophys. {\bf 120} (1996) C317.

\bibitem{rr} R. Ramaty, R. E. Lingenfelter, Astron.  $\&$ Astrophys. Suppl.
 Series {\bf 97} (1993) 127.

\bibitem{cc} E. D. Carlson and L.-S. Tseng, Report HUTP-95/A025
 and hep-ph/9507345.

\bibitem{glast1} P. F. Michelson, SPIE Proc. {\bf 2806} (1996) 31.

\bibitem{osse} W. N. Johnson et al., ApJ. Suppl. {\bf 86} (1993) 693.

\bibitem{carboni} E.g.: U. Amaldi et al.,
 Phys. Lett. {\bf 153B} (1985) 444;
 S. Orito et al., Phys. Rev. Lett. {\bf 63} (1989) 597; M. V. Akopyan et al.,
 Phys. Lett. {\bf B272} (1991) 443;
 S. Asai et al., Phys. Rev. Lett. {\bf 66} (1991) 2440;
 T. Mitsui et al., Phys. Rev. Lett. {\bf 70} (1993) 2265;
 T. Maeno et al., Phys. Lett. {\bf B351} (1995) 574;
 M. Skalsey, R. S. Conti Phys. Rev. {\bf A55} (1997) 984.

\bibitem{surface} M. Ambrosio et al., ApJ. {\bf 464} (1996) 954,
 {\it ibid} {\bf 415} (1993) L147, Phys. Rev. {\bf D47} (1993) 2675;
 A. Borione et al., Phys. Rev. {\bf D49} (1994) 1171.

\bibitem{ambrosio} M. Ambrosio et al., The MACRO collaboration, hep-ex/9807006
 (7. 7. 1998).

\bibitem{BP} B. Paczy\'nski, astro-ph/9706232 (23. 6. 1997), and,
 4th Huntsville GRB

 Symposium/Huntsville, Sept. 1997,
 s. also astro-ph/9712123 (8. 12. 1997).

\bibitem{dar} A. Dar, ApJ. {\bf 500} (1998) L93.

\bibitem{nomad1} S. N. Gninenko and N. V. Krasnikov, Phys. Lett. {\bf B427} (1998) 307;

J. Altegoer et al., The NOMAD collaboration,
 Phys. Lett. {\bf B428} (1998) 197.

\bibitem{egret1} H. A. Mayer-Hasselwander et al., Astron. $\&$ Astrophys.
 {\bf 335} (1998) 161;

 s. also D. H. Hartmann et al., astro-ph/9709029 (3.9.1997).

\bibitem{BP0} B. Paczy\'nski, ApJ. {\bf 304} (1986) 1.

\bibitem{ra} G. G. Raffelt, Stars as Laboratories for Fundamental Physics,

 The University of Chicago Press, Chicago $\&$ London (1996) p.185.

\bibitem{magnetars} R. C. Duncan and C. Thompson, ApJ. {\bf 392} (1992) L9.

\bibitem{morris} D. E. Morris, Phys. Rev. {\bf D34} (1986) 843.

This work considers X-ray emission
($E_{\gamma}\approx 50~$keV) from the magnetosphere of a pulsar, if
{\sf \it axions} are thermally created in its core (s. also ref. \cite{ra}). 

\end{thebibliography}
\end{document}